\documentclass[12pt,preprint]{aastex}
\usepackage{natbib}
\usepackage{url}
\usepackage{amsmath}
\usepackage{multirow}
\usepackage{color}
\usepackage{ulem}

\shorttitle{MAGNETIC HELICITY AND CMES}
\shortauthors{PARK et al.}

\begin{document}

\title{STUDY OF MAGNETIC HELICITY INJECTION IN THE ACTIVE REGION NOAA 9236 PRODUCING MULTIPLE FLARE-ASSOCIATED CME EVENTS}

\author{SUNG-HONG PARK\altaffilmark{1}, KANYA KUSANO\altaffilmark{2,3}, KYUNG-SUK CHO\altaffilmark{1}, JONGCHUL CHAE\altaffilmark{4}, SU-CHAN BONG\altaffilmark{1}, PANKAJ KUMAR\altaffilmark{1}, SO-YOUNG PARK\altaffilmark{4}, YEON-HAN KIM\altaffilmark{1}, YOUNG-DEUK PARK\altaffilmark{1}}

\altaffiltext{1}{Korea Astronomy and Space Science Institute, Daejeon 305-348, Republic of Korea; freemler@kasi.re.kr}
\altaffiltext{2}{Solar Terrestrial Environment Laboratory, Nagoya University, Nagoya 464-8601, Japan}
\altaffiltext{3}{Japan Agency for Marine-Earth Science and Technology (JAMSTEC), Yokohama, Kanagawa 236-0001, Japan}
\altaffiltext{4}{Astronomy Program, Department of Physics and Astronomy, Seoul National University, Seoul, 151-742, Republic of Korea}

\begin{abstract}
To better understand a preferred magnetic field configuration and its evolution during Coronal Mass Ejection events, we investigated the spatial and temporal evolution of photospheric magnetic fields in the active region NOAA 9236 that produced eight flare-associated CMEs during the time period of 2000 November 23--26. The time variations of the total magnetic helicity injection rate and the total unsigned magnetic flux are determined and examined not only in the entire active region but also in some local regions such as the main sunspots and the CME-associated flaring regions using $SOHO$/MDI magnetogram data. As a result, we found that: (1) in the sunspots, a large amount of postive (right-handed) magnetic helicity was injected during most of the examined time period, (2) in the flare region, there was a continuous injection of negative (left-handed) magnetic helicity during the entire period, accompanied by a large increase of the unsigned magnetic flux, and (3) the flaring regions were mainly composed of emerging bipoles of magnetic fragments in which magnetic field lines have substantially favorable conditions for making reconnection with large-scale, overlying, and oppositely directed magnetic field lines connecting the main sunspots. These observational findings can also be well explained by some MHD numerical simulations for CME initiation (e.g., reconnection-favored emerging flux models). We therefore conclude that reconnection-favored magnetic fields in the flaring emerging flux regions play a crucial role in producing the multiple flare-associated CMEs in NOAA 9236.
\end{abstract}


\section{Introduction}
\label{sec1}
Coronal mass ejections (CMEs) are large-scale transient eruptions of magnetized plasma from the solar corona that propagate outward into interplanetary space. Their statistical, physical, and morphological properties have been well studied using satellite- and ground-based observational data, especially by using white-light coronagraph data from the Large Angle and Spectrometric Coronagraph \citep[LASCO,][]{Brueckner:1995} aboard the $SOHO$ spacecraft \citep[e.g.,][]{Yashiro:2004,Gopalswamy:2006,Gopalswamy:2009}. On the other hand, in an effort to understand the initiation mechanism of CMEs, a number of numerical simulations have been carried out with different initial magnetic configurations and prescribed motions at the photosphere, i.e., (1) a reconnection-favored emerging magnetic flux model \citep[e.g.,][]{Chen:2000,Archontis:2008,Kusano:2012}, (2) a flux cancellation model \citep[e.g.,][]{vanBallegooijen:1989,Linker:2001}, (3) a breakout model \citep[e.g.,][]{Antiochos:1999}, (4) ideal MHD instabilities such as kink instability \citep[e.g.,][]{Hood:1981,Fan:2004,Torok:2005} and torus instability \citep[e.g.,][]{Bateman:1978,Titov:1999,Kliem:2006}, (5) shear motion of footpoints of magnetic field lines \citep[e.g.,][]{Mikic:1988,Kusano:2003b,Kusano:2004,Manchester:2008}. Meanwhile, observational studies have demonstrated the common features of magnetic fields and plasma flows in the CME-producing active regions: (1) highly sheared magnetic fields \citep[e.g.,][]{Falconer:2002,Falconer:2006} and strong shear flows \citep[e.g.,][]{Yang:2004,Tan:2009,Liu:2012} were often observed in the vicinity of the photospheric magnetic polarity inversion lines of CME-producing active regions and (2) newly emerging magnetic flux \citep[e.g.,][]{Feynman:1995,Jing:2004,Wang:2004} and sunspot rotations \citep[e.g.,][]{Kumar:2010,Yan:2012,Torok:2013} were reported in some of CME-productive active regions as possible mechanisms for CME initiation. However, the magnetic structure and the plasma flow in the CME-prodcuing active regions as well as the triggering mechanisms are still yet to be determined.

It is believed that CMEs occur in solar active regions with complex magnetic structures, whose complexity and non-potentiality can be measured by magnetic helicity quantitatively, in terms of twists, kinks and inter-linkages of the magnetic field lines \citep{Berger:1984,Pevtsov:2008,Demoulin:2009}. Magnetic helicity studies therefore have been carried out to understand magnetic field configurations and pre-eruption conditions in eruptive solar active regions. \citet{Romano:2003} studied a filament in the active region NOAA 8375, and found a steady injection of magnetic helicity through the photosphere over a period of $\sim$28 hours involving four eruptive events in the filament. It was also found that major flares were preceded by a large amount of photospheric helicity injection in their source active regions over a period of a few days \citep[e.g.,][]{LaBonte:2007,Park:2008,Park:2010a}. In particular, there were some observational reports that the injection of oppositely signed magnetic helicity into a pre-existing coronal field might be a crucial ingredient for triggering eruptions of active-region filaments, flares, and CMEs: e.g., \citet{Park:2010b} investigated coronal magnetic helicity in the flare-CME-productive active region NOAA 10930, and found that there was not only a large increase of negative magnetic helicity of $-$3.2$\times$10$^{43}\,$Mx$^2$ in the active region corona over $\sim$1.5 days prior to the X3.4 flare and its associated halo CME on 2006 December 13 but also a significant injection of positive magnetic helicity through the photosphere around the flaring magnetic polarity inversion line of the active region \citep[for other observational reports, refer also to][]{Kusano:2003a,Yokoyama:2003,Romano:2011a,Jing:2012,Kumar:2013}. 

Recently, by investigating photospheric magnetic helicity injection in 28 active regions producing 47 CMEs, \citet{Park:2012} found that the 47 CMEs under investigation can be categorized into two different groups by the evolution of helicity injection in their source active regions: (1) a monotonically increasing pattern with one sign of helicity (Group A; 30 CMEs) and (2) a pattern of significant helicity injection followed by its sign reversal (Group B; 17 CMEs). In addition, it was found that there was a difference in CME speed, acceleration, and occurrence frequency between Group A and Group B. The main difference between these two CME groups in the temporal variation of helicity injection is that the helicity sign reversal phase appears in Group B, but not in Group A. In the study of \citet{Park:2012}, however, were not examined (1) where oppositely signed helicity is significantly injected during the sign reversal phase and (2) how its magnitude and distribution evolve with time. In this study, we therefore investigate carefully the spatial distribution and temporal variation of magnetic helicity flux density (i.e., magnetic helicity injection per unit area per unit time) on the photosphere of the very eruptive active region NOAA 9236. This active region clearly showed the sign reversal phase and produced eight large CMEs during the sign reversal phase. We expect that this study will develop a better understanding of a pre-CME magnetic structure and a trigger mechanism for CME initiation. The rest of this paper is organized as follows: data selection and analysis for the calculation of magnetic helicity flux density in NOAA 9236 are explained in Section~\ref{sec2}. In Section~\ref{sec3}, we present the observational findings that: (1) all of eight CME-associated flares occurred in newly emerging magnetic flux regions in the vicinity of the main positive-polarity sunspot of NOAA 9236 and (2) magnetic fields in the CME-associated flaring regions have favorable conditions for making reconnection with large-scale and overlying magnetic fields in the main sunspot regions of positive and negative magnetic polarities. Finally, the results are summarized and discussed in Section~\ref{sec4}.

\section{Data Selection and Analysis}
\label{sec2}

\subsection{Overview of Active Region NOAA 9236}
\label{sec2:sub1}
The active region NOAA 9236 appeared on the east limb of the solar disk on 2000 November 18, and it was continuously observed during the time interval of its entire disk passage by the $SOHO$ spacecraft. At an early stage of the development of NOAA 9236 (i.e., 2000 November 18--22), a group of sunspots in the active region showed a general $\beta$ magnetic configuration of two main sunspots of positive (preceding) and negative (following) magnetic polarities with a simple and distinct division between the two polarities. From 2000 November 23, it then changed to a $\beta\gamma$ configuration: i.e., a number of newly emerging small sunspot groups with complex polarities appeared around the main positive-polarity sunspot. 

There have been some studies on the properties of (1) flare-CME events that occurred in the active region NOAA 9236 and (2) magnetic fields on the photosphere. \citet{Nitta:2001} examined 6 recurrent halo CMEs and their associated flares in NOAA 9236, and reported that the flares were not long-decay events \citep[LDE,][]{Kahler:1977} in terms of soft X-ray light curves and morphologies even though they produced soft X-ray plasma ejections during their impulsive phases. They also found that soft and hard X-ray emissions from the flares were seen in and around the preceding positive-polarity sunspot. \citet{Zhang:2002} studied temporal variation of magnetic flux of small moving magnetic features (MMFs) in the vicinity of the main positive sunspot in NOAA 9236 by tracking 452 MMFs from birth to death, and found that magnetic flux emergence and disappearance rates in the form of MMFs are 2.3$\times$10$^{20}\,$Mx$\,$hr$^{-1}$ and 1.9$\times$10$^{20}\,$Mx$\,$hr$^{-1}$, respectively, during the period of 2000 November 23--24. \citet{Takasaki:2004} analyzed multi-wavelength observational data of three X-class homologous flares in NOAA 9236 that occurred on 2000 November 24. As a result, they found a good correlation between the separation velocity of the two ribbons and the time derivative of the soft X-ray flux for each of the homologous flares, which can be explained by the magnetic energy release rate estimated from magnetic reconnection models for solar flares \citep[e.g.,][]{Shibata:1999,Isobe:2002}.

In this study, we only took into account the time span of $SOHO$/Michelson Doppler Imager \citep[MDI,][]{Scherrer:1995} observations from 2000 November 23, 11:12 UT to 2000 November 26, 20:47 UT when the active region was located near the center of the apparent solar disk. This was to reduce the uncertainty of determined magnetic helicity flux density in NOAA 9236 produced in the derivation of magnetic fields perpendicular to the photosphere from MDI measurements of line-of-sight magnetic fields. During the time interval under consideration, eight large CMEs, including six halo CMEs, occurred in the active region accompanied with flares (see Table~\ref{sec2:tab1} for the detailed information of the multiple flare-associated CME events). Note that the CMEs were identified in the originating active region NOAA 9236 by carefully inspecting the locations of soft X-ray and extreme-ultraviolet (EUV) brightenings of their associated flares, observed by the $Yohkoh$/Soft X-ray Telescope \citep[SXT,][]{Tsuneta:1991} and $SOHO$/Extreme-ultraviolet Imaging Telescope \citep[EIT,][]{Delaboudini:1995}, respectively. 

\subsection{Calculation of Photospheric Magnetic Helicity Injection}
\label{sec2:sub2}
By helicity, we refer to the {\it relative} magnetic helicity $H$ in the rest of this paper, i.e., the gauge-invariant amount of magnetic helicity derived by \citet{Finn:1985}:
\begin{equation}
\label{sec2:sub2:equ1}
H=\int_{V} ({\textit{\textbf{A}}}+{\textit{\textbf{A}}}_{p}) \cdot ({\textit{\textbf{B}}}-{\textit{\textbf{P}}})\,dV,
\end{equation}
where $\textit{\textbf{P}}$ is the potential field having the same normal component as the magnetic field $\textit{\textbf{B}}$ on the boundary surface of the volume $V$ of a given magnetic field system. $\textit{\textbf{A}}$ and $\textit{\textbf{A}}_{p}$ are the vector potentials for $\textit{\textbf{B}}$ and $\textit{\textbf{P}}$, respectively.

Helicity injected through the photospheric surface of a solar active region can be determined by taking into account helicity flux density $G_{\theta}$ (i.e., helicity injection per unit area per unit time). We calculated $G_{\theta}(\textit{\textbf{x}},t)$ at a position $\textit{\textbf{x}}$ on the entire surface $S_{E}$ of NOAA 9236 and at a specific time $t$ using the well-recognized formula of $G_{\theta}$ proposed by \citet{Pariat:2005} with the help of the numerical calculation method developed by \citet{Chae:2007}: 
\begin{equation}
\label{sec2:sub2:equ2}
G_{\theta}(\textit{\textbf{x}},t)=-\frac{B_{n}}{2\pi} \int_{S_{E}} \left(\frac{\textit{\textbf{x}}-\textit{\textbf{x}}\,'}{|\textit{\textbf{x}}-\textit{\textbf{x}}\,'|^{2}} \times [\textit{\textbf{u}}-\textit{\textbf{u}}'] \right)_{n} B_{n}' \,dS',
\end{equation}
where the subscript $n$ indicates the component perpendicular to $S_{E}$. $B_{n}$ is the magnetic field component perpendicular to $S_{E}$, and it is derived from the line-of-sight component $B_{l}$ of $SOHO$/MDI magnetogram data with a spatial resolution of 2$\arcsec$ per pixel by assuming that the horizontal magnetic field component to $S_{E}$ is negligible compared to $B_{l}$ (i.e., $B_{n}$ = $B_{l}$/$\cos{\psi}$ where $\psi$ is the heliocentric angle). $\textit{\textbf{u}}$ is the velocity of the apparent horizontal motion of magnetic field line footpoints at the surface $S_{E}$ determined by the differential affine velocity estimator (DAVE) method \citep{Schuck:2006}. In applying the DAVE, we set the full width at half maximum of a localizing window and the time interval between two successive aligned images as 10$\arcsec$ and 96 minutes, respectively. Refer to the study of \citet{Chae:2007} for the details of the $G_{\theta}$ calculation. Note also that many recent studies have used $G_{\theta}$ for calculating helicity injection rate in solar active regions \citep[e.g.,][]{Park:2010b,Smyrli:2010,Romano:2011a,Romano:2011b,Zuccarello:2011,Park:2012}.  

\begin{figure}
\begin{center}\epsscale{0.65}
\plotone{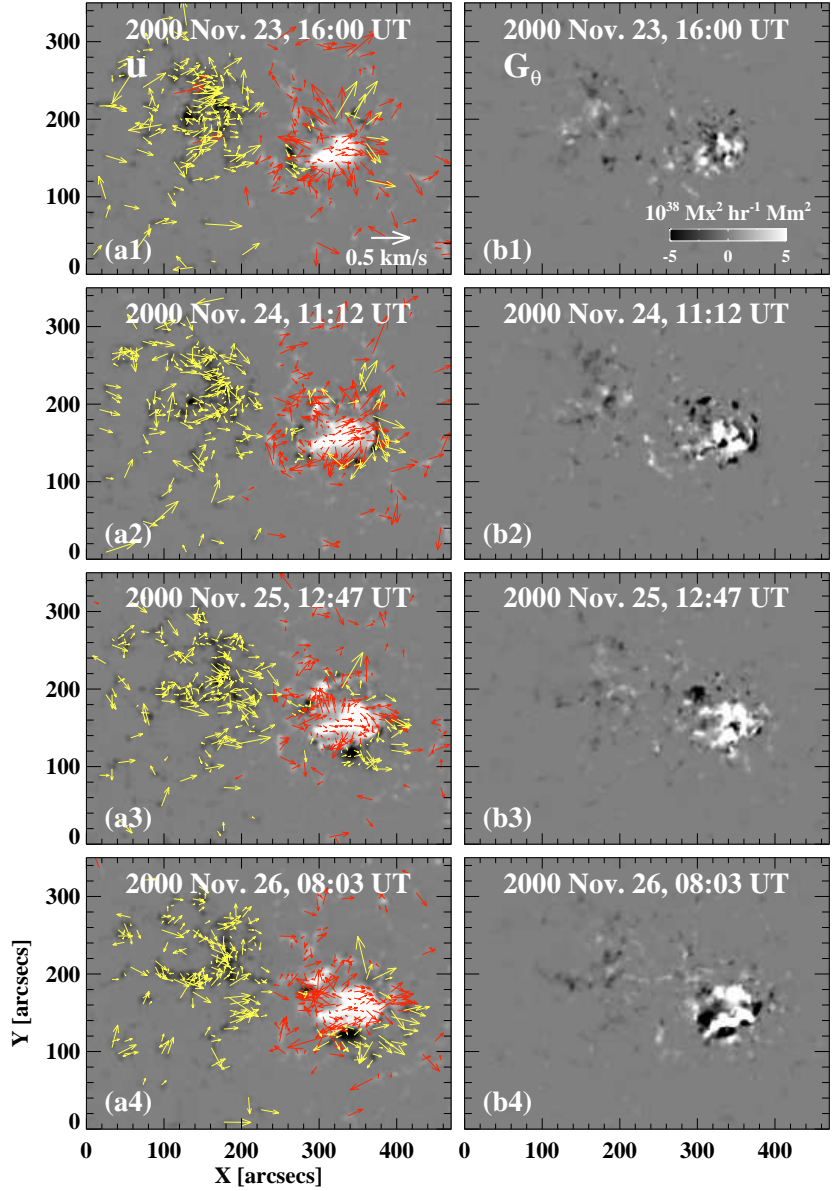}
\caption{Active region NOAA 9236 during the time span of 2000 November 23--26. Left: magnetic footpoint velocity $\textit{\textbf{u}}$ is superposed on a gray-scale image of the normal magnetic field component $B_{n}$ as red/yellow arrows on positive/negative $B_{n}$. Right: positive/negative values of helicity flux density $G_{\theta}$, corresponding to right-handed/left-handed helicity flux densities, are displayed as white/black tones. Note that the saturation level of $G_{\theta}$ is set as $\pm$5$\times$10$^{38}$ Mx$^2$ hr$^{-1}$ Mm$^{-2}$ for the purpose of display visibility.}
\label{sec2:sub2:fig1}
\end{center}
\end{figure}

Figure~\ref{sec2:sub2:fig1} presents some of $B_{n}$, $\textit{\textbf{u}}$, and $G_{\theta}$ maps of the active region NOAA 9236 during the time span of 2000 November 23--26: i.e., the velocity $\textit{\textbf{u}}$ is superposed on a gray-scale image of $B_{n}$ as red and yellow arrows on positive and negative $B_{n}$, respectively, in the left panels, and $G_{\theta}$ is shown by a gray-scale image in the right panels. The velocity maps clearly show that there are steady outward motions of MMFs from the main positive sunspot with an average speed of 0.15$\,$km$\,$s$^{-1}$ during the observation time. It is also shown in $G_{\theta}$ maps that there is a continuous injection of positive helicity inside the main positive sunspot, while there is a remarkable injection of negative helicity in MMF regions around the positive sunspot. The details of temporal variations and spatial distributions of $G_{\theta}$ in NOAA 9236 are described in Section~\ref{sec3}. 
 
We calculated the total helicity injection rate $\dot{H}$ through a local region $S_{L}$ of NOAA 9236 by integrating $G_{\theta}(\textit{\textbf{x}},t)$ in the area of $S_{L}$:
\begin{equation}
\label{sec2:sub2:equ3}
\dot{H}(t)=\int_{S_{L}} G_{\theta}(\textit{\textbf{x}},t)\,dS_{L}.
\end{equation}
Then, the accumulated amount of the total helicity injection $\Delta H$ through $S_{L}$ was determined by integrating $\dot{H}$ with respect to time:
\begin{equation}
\label{sec2:sub2:equ4}
\Delta H(t) = \int_{t_0}^{t} \dot{H}(t')\,dt',
\end{equation}
where $t_0$ is the start time of the MDI magnetogram data set of the active region. Positive and negative values of $\dot{H}$ indicate that a dominant sign of helicity injected through $S_{L}$ per unit time is right-handed and left-handed senses, respectively. As a reference parameter, we also derived the total unsigned magnetic flux $\Phi$ at $S_{L}$:
\begin{equation}
\label{sec2:sub2:equ5}
\Phi(t) = \int_{S_{L}} |B_{n}(\textit{\textbf{x}},t)|\,dS_{L}.
\end{equation}

It is important to check whether the calculation of $\dot{H}$ is sensitively affected by the per-pixel noise ($\sim$30$\,$G) in MDI magnetograms. We therefore estimated the uncertainty of $\dot{H}$ that is originated from measurement uncertainty of MDI magnetograms as follows. First, pseudo-random noises which have normal distribution with the standard deviation of 30 G were added to each magnetogram. Then $\dot{H}$ through the entire region $S_{E}$ of NOAA 9236 was calculated. The calculation of $\dot{H}$ was repeated 10 times with different sets of errors to calculate the standard deviations of $\dot{H}$. Finally, we calculated the average of the standard deviations during the measurement period of $\dot{H}$, and considered it as the uncertainty of $\dot{H}$. It was found that the uncertainty is 6.8$\times$10$^{39}\,$Mx$^{2}\,$hr$^{-1}$, i.e., 3$\%$ of the average of $\dot{H}$. This indicates that the MDI noise does not significantly affect the helicity calculation.

\section{Results}
\label{sec3}
\begin{figure}
\begin{center}\epsscale{0.6}
\plotone{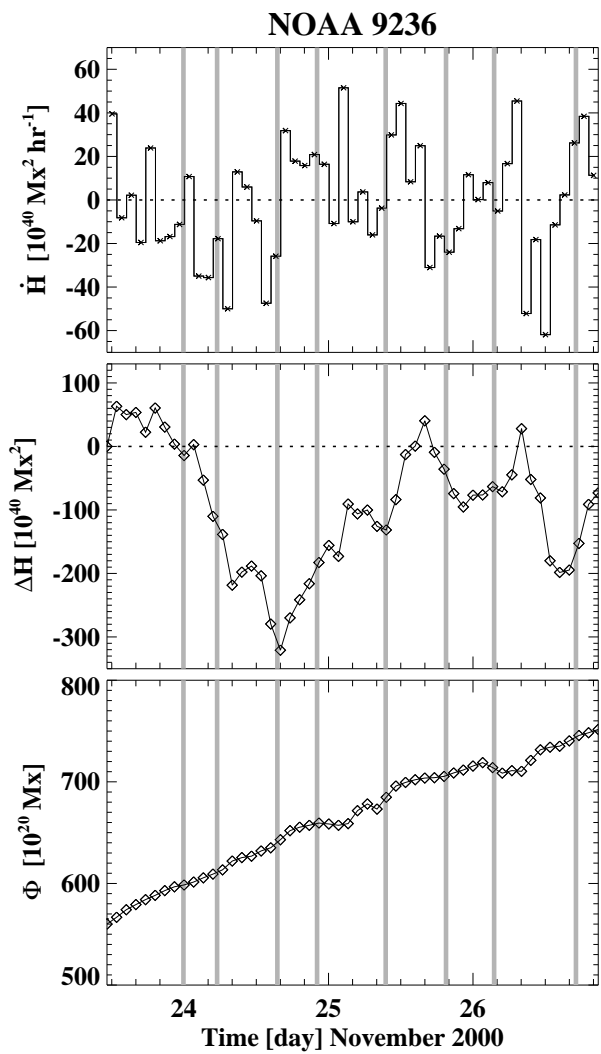}
\caption{Temporal variations of the total helicity injection rate $\dot{H}$ (top), the accumulated amount of the total helicity injection $\Delta H$ (middle), and the total unsigned magnetic flux $\Phi$ (bottom) measured in the entire photospheric surface of the active region. During the measurement period, both positive and negative magnetic helicities were concurrently and continuously injected with a significant and continuous increase of the magnetic flux. The vertical lines indicate the first appearance times of the eight CMEs in the LASCO/C2 field-of-view.}
\label{sec3:fig1}
\end{center}
\end{figure}

We first checked how helicity injection and magnetic flux, respectively, through and in the entire photospheric surface of the active region evolve in time. Figure~\ref{sec3:fig1} presents the time profiles of $\dot{H}$ (top), $\Delta H$ (middle), and $\Phi$ (bottom) calculated for $S_{E}$ of NOAA 9236. Note that the vertical lines in Figures~\ref{sec3:fig1},~\ref{sec3:fig4}, and~\ref{sec3:fig5} represent the first appearance of each of the eight CMEs in the field of view (FOV) of LASCO/C2. The temporal variations of $\dot{H}$ and $\Delta H$ show that a relatively large amount of negative helicity started to be steadily injected through $S_{E}$ for the first 28 hours (November 23, 12:00 UT -- November 24, 16:00 UT) of the helicity measurement with an average helicity injection rate of $-$11$\times$10$^{40}\,$Mx$^{2}\,$hr$^{-1}$. And then the injection of a significant amount of oppositely signed (positive) helicity through $S_{E}$ took place for the next 24 hours (November 24, 16:00 UT -- November 25, 16:00 UT) with an average rate of 15$\times$10$^{40}\,$Mx$^{2}\,$hr$^{-1}$. Afterwards, $\dot{H}$ through $S_{E}$ changed its sign every 6 hours until the end of the measurement (November 25, 16:00 UT -- November 26, 20:00 UT). This result indicates that both positive and negative helicities were concurrently, significantly, and continuously injected through $S_{E}$ during 80 hours of the entire measurement period. On the other hand, $\Phi$ in $S_{E}$ kept increasing from 5.6$\times$10$^{22}\,$Mx to 7.5$\times$10$^{22}\,$Mx with an average rate of 6$\times$10$^{21}\,$Mx$\,$hr$^{-1}$. The increase of $\Phi$ is mainly due to newly emerging magnetic flux around the main positive sunspot. Note that the characteristic evolution pattern of significant helicity injection at the photosphere followed by the sign reversal of injected helicity has been well reported by previous studies of photospheric helicity injection in flare-CME-productive active regions \citep[e.g.,][]{Park:2010a,Park:2012}. However, the understanding of the occurrence of flare-assocaited CMEs requires further study on the spatial distribution and temporal evolution of the positive and negative components of the injected helicity.

\begin{figure}
\begin{center}\epsscale{0.65}
\plotone{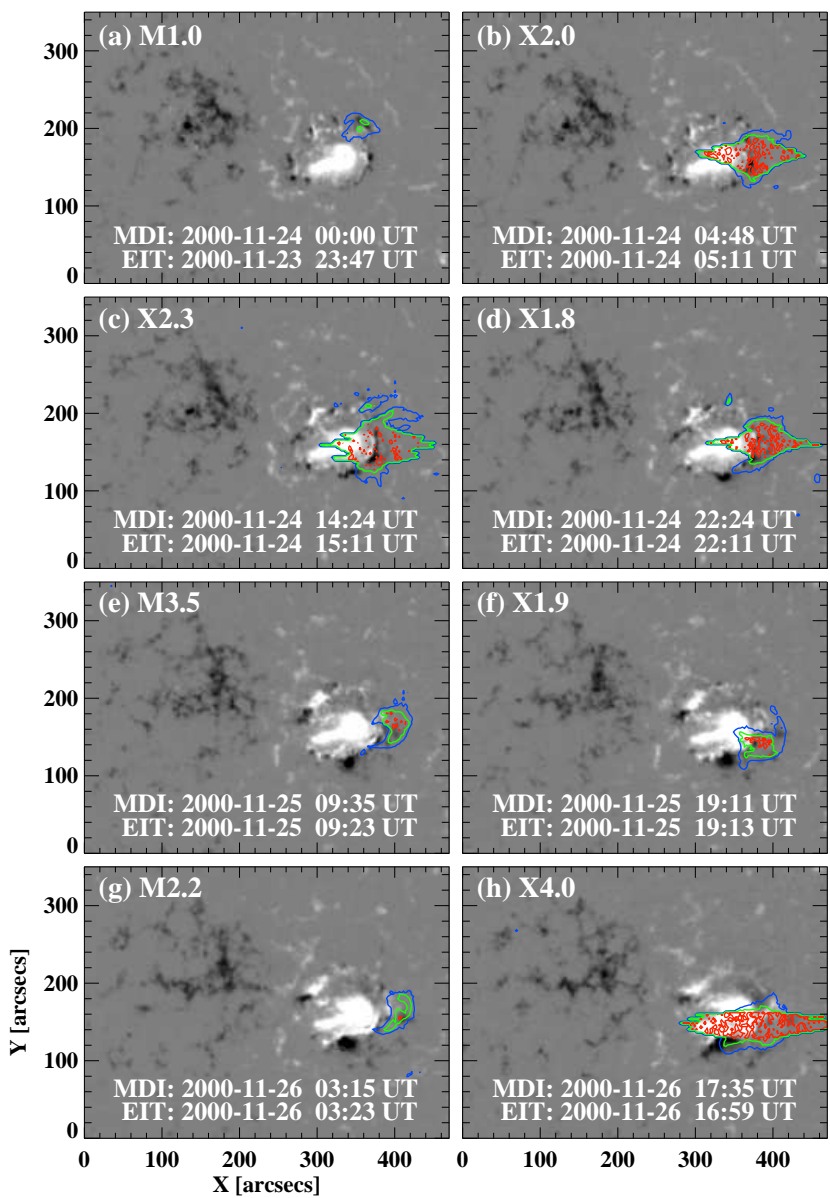}
\caption{$B_{n}$ maps (gray-scale images) aligned with EIT 195$\,${\AA} maps (contours) at the times of eight CME-associated flares. The blue, green, and red contour lines represent the EIT 195$\,${\AA} emission corresponding to (2, 6, and 12)$\times$10$^{3}\,$DN per pixel, respectively. DN stands for Data Number, a generic counts of digitized camera signal. All of the eight flares occurred in newly emerging magnetic flux regions in the vicinity of the main positive-polarity sunspot of NOAA 9236.}
\label{sec3:fig2}
\end{center}
\end{figure}

\begin{figure}
\begin{center}\epsscale{0.39}
\plotone{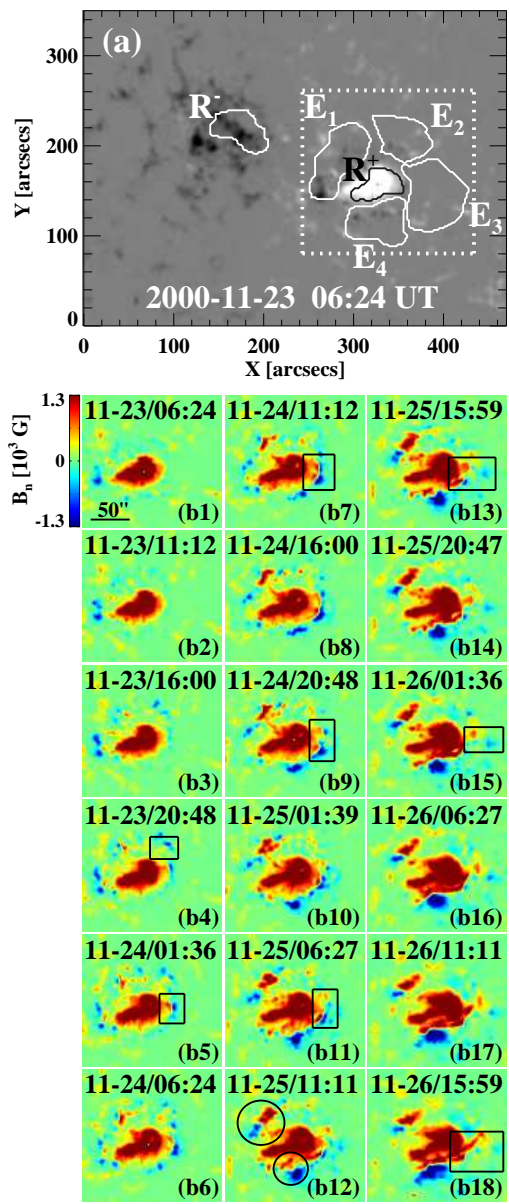}
\caption{Panel (a): some local regions are marked on a map of $B_{n}$ of NOAA 9236. R$^{+}$ and R$^{-}$ represent the main positive and negative-polarity sunspot regions, respectively. E$_{1}-$E$_{4}$ indicate emerging flux regions around R$^{+}$. Note that E$_{2}$ and E$_{3}$ are flaring regions. Panel (b): a time series of $B_{n}$ maps of the local area, marked with the dotted-line box in the panel (a), is presented. Emerging bipolar magnetic regions around the occurrence times and locations of the eight CME-associated flares under consideration are marked with the rectangular boxes in their corresponding panels. The circles in the panel b12 indicate two bipolar pairs of emerging magnetic fragments located in E$_{1}$ and E$_{4}$, respectively.}
\label{sec3:fig3}
\end{center}
\end{figure}

\begin{figure}
\begin{center}\epsscale{0.7}
\plotone{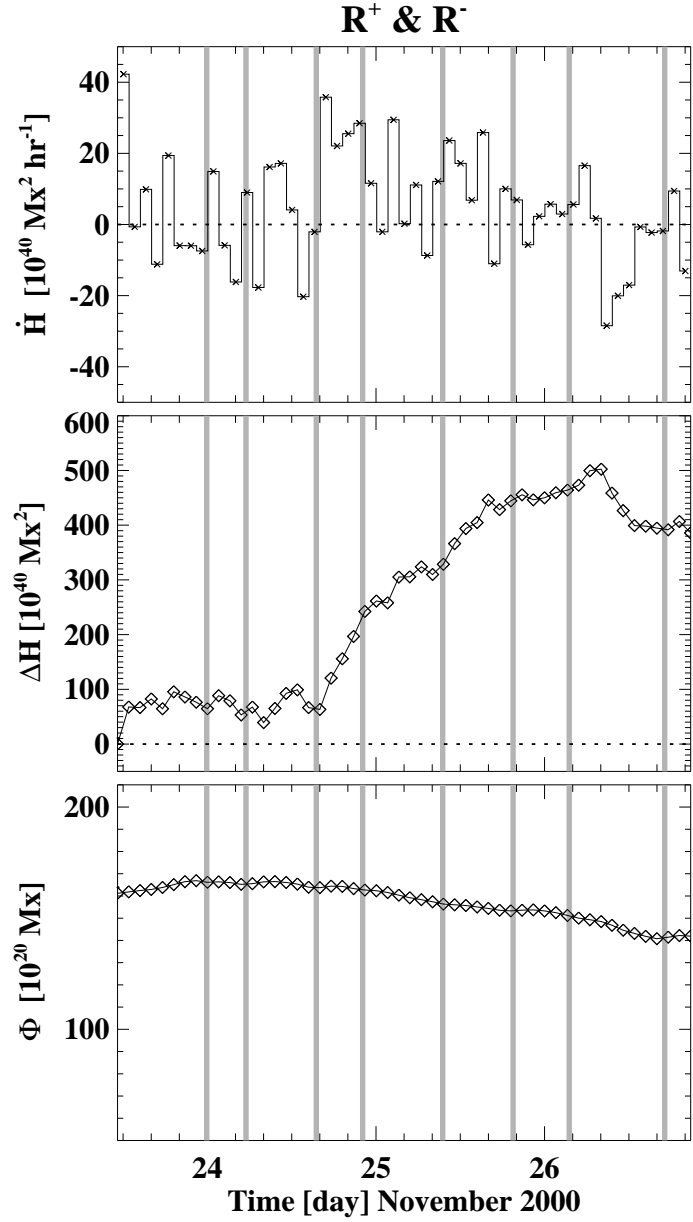}
\caption{Same as in Figure~\ref{sec3:fig1}, but for the main positive and negative sunspot regions, R$^{+}$ and R$^{-}$, marked in Figure~\ref{sec3:fig3}{a}. A large amount of positive helicity injection through the photospheric surfaces of R$^{+}$ and R$^{-}$ took place during most of the measurement time. The total unsigned magnetic flux $\Phi$, however, changed little.}
\label{sec3:fig4}
\end{center}
\end{figure}

\begin{figure}
\begin{center}\epsscale{1}
\plotone{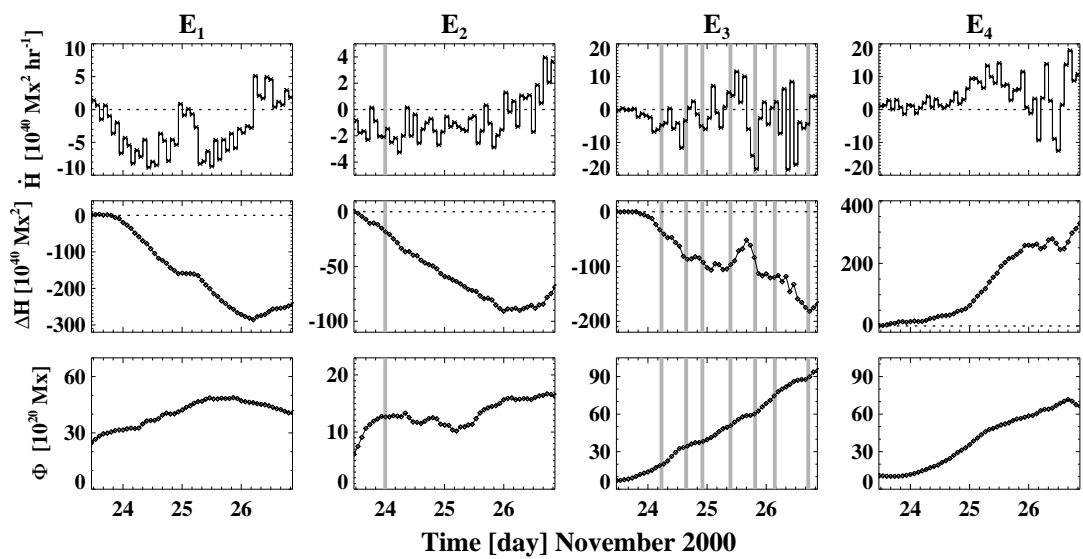}
\caption{Same as in Figure~\ref{sec3:fig1}, but for the emerging flux regions, E$_{1}-$E$_{4}$, marked in Figure~\ref{sec3:fig3}{a}. There was a continuous injection with one dominant sign of helicity in each of the regions (i.e., negative sign in E$_{1}$, E$_{2}$, and E$_{3}$; positive sign in E$_{4}$) during the entire measurement period, accompanied by an increase of the total unsigned magnetic flux $\Phi$.}
\label{sec3:fig5}
\end{center}
\end{figure}

We therefore aligned MDI magnetograms of NOAA 9236 with EIT 195$\,${\AA} images to identify local regions where intense EUV emission occurred from the CME-associated flares. Figure~\ref{sec3:fig2} shows the active region MDI maps (gray-scale images) with the aligned EIT contours at the times of the eight CME-associated flares. As shown in Figure~\ref{sec3:fig2}, all the CME-associated flares occurred in the vicinity of the main positive sunspot: i.e., the M1.0 flare on 2000 November 23 and the following seven flares occurred in the northern and western neighboring regions of the main positive sunspot, respectively. For a more detailed understanding of morphological characteristics of magnetic fields in the flaring regions, we selected a local area enclosing emerging flux regions around the main positive sunspot (see the dotted-line box in Figure~\ref{sec3:fig3}{a}) and examined a time series of $B_{n}$ maps of the local area for the period of 2000 November 23--26 with a time cadence of 5 hours, as shown in Figure~\ref{sec3:fig3}{b}. In Figure~\ref{sec3:fig3}{a}, R$^{+}$ and R$^{-}$ represent the main positive and negative-polarity sunspot regions. E$_{1}-$E$_{4}$ indicate the emerging flux regions which are located in the eastern, northern, western, and southern part of the positive sunspot, respectively: E$_{2}$ and E$_{3}$ regions correspond to the flaring regions, while E$_{1}$ and E$_{4}$ represent non-flaring regions. Figure~\ref{sec3:fig3}{b} shows that the emerging flux regions are mainly composed of small-scale bipolar magnetic fragments. We especially marked some bipoles of emerging magnetic fragments   which appear around the times and in the locations of the eight flares with rectangular boxes in the panels of Figure~\ref{sec3:fig3}{b}. The bipolar magnetic fragment in E$_{2}$ (as indicated by the box at the panel b4 of Figure~\ref{sec3:fig3}{b}) first appeared at 16:00 UT on 2000 November 23, developed over a few hours, and gradually disappeared after the M1.0 flare at 23:18 UT on 2000 November 23. In the multiple-flare-productive region E$_{3}$, we also found relatively larger and more elongated bipolar fragments compared to that in E$_{2}$ during the flare occurrence times of 2000 November 24-26. It is noteworthy that \citet{Wang:2002} reported elongated magnetic structures in E$_{3}$, which is associated with the X1.8 flare on 2000 November 24, by examining a high-resolution line-of-sight magnetogram obtained with the Big Bear Solar Observatory's (BBSO) Digital Vector Magnetograph system \citep[DVMG,][]{Wang:1998}. In addition, two bipolar pairs of emerging magnetic fragments located in E$_{1}$ and E$_{4}$, respectively, are marked with the circles in the panel b12 of Figure~\ref{sec3:fig3}{b}.   

Next, we determined the temporal evolution of $\dot{H}$, $\Delta H$, and $\Phi$ in the local region of the main positive and negative sunspots (see Figure~\ref{sec3:fig4}), taking account of Equations~\ref{sec2:sub2:equ3},~\ref{sec2:sub2:equ4}, and~\ref{sec2:sub2:equ5} with the integrated area of R$^{+}$ and R$^{-}$ combined.  In Figure~\ref{sec3:fig4}, the time variations of $\dot{H}$ and $\Delta H$ show the following evolution patterns: (1) a phase of nearly constant $\Delta H$ of 80$\times$10$^{40}\,$Mx$^{2}$ lasted for the first 28 hours (November 23, 12:00 UT -- November 24, 16:00 UT) after an abrupt and large injection of positive helicity at the beginning of the helicity measurement, (2) $\Delta H$ kept on significantly and monotonically increasing up to 500$\times$10$^{40}\,$Mx$^{2}$ at an average rate of 12.5$\times$10$^{40}\,$Mx$^{2}\,$hr$^{-1}$ over the next 40 hours (November 24, 16:00 UT -- November 26, 8:00 UT), and (3) it decreased by 100$\times$10$^{40}\,$Mx$^{2}$ for the next 4.5 hours and remained in a phase of its constant value of 400$\times$10$^{40}\,$Mx$^{2}$ until the end of the measurement. On the other hand, in the case of $\Phi$, it changed little for the entire measurement period with only a slight decrease by $\sim$10$\%$ from 160$\times$10$^{20}\,$Mx to 140$\times$10$^{20}\,$Mx. 

Finally, as shown in Figure~\ref{sec3:fig5}, we also calculated and plotted the time variations of $\dot{H}$, $\Delta H$, and $\Phi$ in each of the emerging flux regions E$_{1}-$E$_{4}$. The following are the main findings from Figure~\ref{sec3:fig5}: (1) the flaring region E$_{2}$ shows a considerable injection of negative helicity (i.e., a decrease of $\Delta H$ of $-$20$\times$10$^{40}\,$Mx$^{2}$) for 4.5 hours before the M1.0 flare and the associated CME on 2000-Nov-23, accompanied by an increase of $\Phi$ from 6$\times$10$^{20}\,$Mx to 13$\times$10$^{20}\,$Mx with an average rate of 15.6$\times$10$^{19}\,$Mx$\,$hr$^{-1}$, (2) the multiple flaring region E$_{3}$ also presents a similar pattern in the temporal variation of $\Delta H$ during the three days of seven CME events: i.e., a significant amount of negative helicity was injected through the region E$_{3}$ over 72 hours so that $\Delta H$ reached $-$180$\times$10$^{40}\,$Mx$^{2}$ with an average rate of $-$2.5$\times$10$^{40}\,$Mx$^{2}\,$hr$^{-1}$. In addition, there was a large increase of $\Phi$ from 10$\times$10$^{20}\,$Mx to 100$\times$10$^{20}\,$Mx with an average rate of 10.9$\times$10$^{19}\,$Mx$\,$hr$^{-1}$ during the entire time period, (3) the non-flaring region E$_{1}$ shows a continuous increase in the magnitude of the negative helicity, however, $\Phi$ only increased from 30$\times$10$^{20}\,$Mx to 40$\times$10$^{20}\,$Mx with a much smaller average rate of 1.8$\times$10$^{19}\,$Mx$\,$hr$^{-1}$ during the entire period, and (4) the non-flaring region E$_{4}$ has a significant amount of injection of positive helicity, which is of the same sign as in the main sunspot, accompanied by an increase of $\Phi$.

We now have the question of why CME-related flares occurred in E$_{2}$ and E$_{3}$, but not in E$_{1}$ and E$_{4}$. In the long-term (a few days) variation of helicity injection, there seems to be no significant difference between the flaring and non-flaring regions because they in general show a similar trend in the time variation of $\Delta H$, i.e., a monotonically increasing trend with one sign of helicity during the entire investigation. Furthermore, the magnitudes of $\dot{H}$, $\Delta H$, and $\Phi$ in the flaring regions are comparable to those in the non-flaring regions. However, comparison of the helicity in the bipolar magnetic fragments of E$_{2}$ and E$_{3}$ and the main sunspots suggests that the magnetic fields in E$_{2}$ and E$_{3}$ (especially in the emerging bipoles outlined by the rectangular boxes in Figure~\ref{sec3:fig3}{b}) have substantially favorable conditions for making reconnection with those in R$^{+}$ and R$^{-}$. The simplified cartoon of Figure~\ref{sec3:fig6} describes a top view of the magnetic field lines connecting R$^{+}$ and R$^{-}$ as well as bipoles of the magnetic fragments in each of E$_{1}-$E$_{4}$. The large-scale kinked structures of a coronal magnetic field line can be inferred from the sign of injected helicity through the photosphere, as shown by previous studies \citep[e.g.,][]{Yamamoto:2005,Luoni:2011}: i.e., coronal fields in general display a forward-S (inverse-S) shape with positive (negative) helicity. In Figure~\ref{sec3:fig6}, the small-scale, under-lying, and inverse-S-shaped field lines (red lines) in the emerging bipoles of magnetic fragments in E$_{2}$ and E$_{3}$ may interact and reconnect with the large-scale, overlying, S-shaped, and oppositely directed field lines (green lines) in R$^{+}$ and R$^{-}$. On the other hand, we speculate that the inferred field lines (blue lines) of an inverse-S and S shape in E$_{1}$ and E$_{4}$, respectively, may pass underneath the field lines (green lines) in R$^{+}$ and R$^{-}$ so that it would be difficult to make a reconnection between the emerging and overlying fields. Another noticeable difference is that the average increasing rate of the total unsigned magnetic flux $\dot{\Phi}_{\mathrm{avg}}$ in E$_{3}$ during the entire time period is relatively larger than in the other regions: i.e., 10.9$\times$10$^{19}\,$Mx$\,$hr$^{-1}$ in the most eruptive region E$_{3}$ where seven out of the eight flares occurred, but (1.8, 1.1, and 7.0)$\times$10$^{19}\,$Mx$\,$hr$^{-1}$ in E$_{1}$, E$_{2}$, and E$_{4}$, respectively. Note that the flaring region E$_{2}$ also shows a large value of $\dot{\Phi}_{\mathrm{avg}}$ of 15.6$\times$10$^{19}\,$Mx$\,$hr$^{-1}$ during 4.5 hours right before the M1.0 flare. In addition, we found that E$_{3}$ shows not only a continuous increase of $\Phi$ during the seven flares but also an impulsive injection of negative helicity a few hours before each of the flares, as shown in Figure~\ref{sec3:fig5}. We therefore conjecture that the continuous emergence of magnetic bipoles with an opposite helicity to the overlying fields may be responsible for the occurrence of the multiple flare events in E$_{3}$.
 
\section{Summary and Conclusion}
\label{sec4}
The detailed investigation of the spatial distribution and temporal variation of magnetic helicity injection in a CME-productive active region has been carried out to better understand a preferred magnetic field configuration and its evolution during CME events. In this study, we determined and examined the time variations of the total helicity injection rate $\dot{H}$, the accumulated amount of the total helicity injection $\Delta H$, and the total unsigned magnetic flux $\Phi$ not only in the entire active region NOAA 9236 but also in some local regions (i.e., the main positive and negative sunspot regions, R$^{+}$ and R$^{-}$, and the emerging flux regions, E$_{1}-$E$_{4}$, around R$^{+}$). Note that eight CMEs and their associated flares occurred in E$_{2}$ and E$_{3}$ during the investigation period (2000 November 23--26). As a result, we found that: (1) a large amount of positive helicity was injected through R$^{+}$ and R$^{-}$ during most of the measurement time, (2) a continuous injection with negative sign of helicity took place in the CME-associated flaring regions E$_{2}$ and E$_{3}$ during the entire measurement period, accompanied by an increase of magnetic flux (especially, with a large increasing rate of $\Phi$ in the most eruptive region E$_{3}$), and (3) E$_{2}$ and E$_{3}$ are mainly composed of emerging bipoles of magnetic fragments in which magnetic field lines have substantially favorable conditions for making reconnection with large-scale, overlying, and oppositely directed magnetic field lines in R$^{+}$ and R$^{-}$.

\begin{figure}
\begin{center}\epsscale{0.8}
\plotone{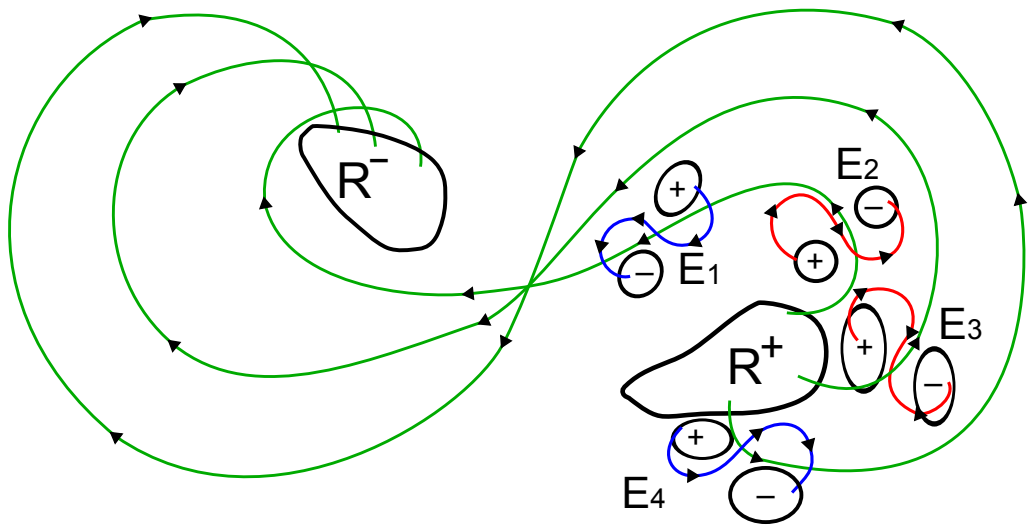}
\caption{This cartoon presents a top view of some of magnetic field lines connecting the main sunspot regions R$^{+}$ and R$^{-}$ as well as a bipolar pair of magnetic fragments in each of the emerging flux regions including both the flaring regions (E$_{2}$ and E$_{3}$) and non-flaring regions (E$_{1}$ and E$_{4}$). The large-scale kinked structures of the field lines are inferred from the signs of injected helicity through the photospheric surfaces in which the field line footpoints anchored. The green lines indicate large-scale, overlying, and S-shaped magnetic field lines of which footpoints anchored in R$^{+}$ and R$^{-}$, while the red lines represent small-scale, under-lying, and inverse-S-shaped field lines in the bipolar pairs of magnetic fragments in E$_{2}$ and E$_{3}$. We can conjecture that the field lines in E$_{2}$ and E$_{3}$ may have a substantially favorable condition for making reconnection with the oppositely directed field lines in R$^{+}$ and R$^{-}$ at the locations where they encounter each other. The blue lines, on the other hand, denote inverse-S-shaped and S-shaped field lines in E$_{1}$ and E$_{4}$, respectively, that may pass underneath the transversely directed field lines in R$^{+}$ and R$^{-}$.}
\label{sec3:fig6}
\end{center}
\end{figure}

We also found that the observational results during the eight flare-associated CMEs events  can be well explained by reconnection-favored emerging magnetic flux models \citep[e.g.,][]{Chen:2000,Archontis:2008,Kusano:2012}. For example, in the two-dimensional CME model of \citet{Chen:2000}, they set up a newly emerging magnetic flux system, of which magnetic fields oppositely directed to and may reconnect with those of a nearby large-scale magnetic flux system, and showed that the large-scale system could be totally destabilized and erupted due to a localized reconnection between the two systems. Although the coronal magnetic field illustrated in Figure~\ref{sec3:fig6} is much more complicated than that in the simple two-dimensional model, it is likely that some internal reconnections between large-scale overlying field lines (R$^{+}$ to R$^{-}$) and small-scale emerging bipolar fields (E$_{2}$ and E$_{3}$) may destabilize the magnetic system and give rise to CMEs. In fact, \citet{Kusano:2012} recently found that the emergence of small-scale flux could trigger large-scale solar eruption even in three-dimensional (3D) systems, if the emerging flux has the magnetic helicity with the sign opposite to the overlying field or the magnetic polarity of the emerging flux is oppositely directed to the large-scale field. The small-scale magnetic fluxes in regions E$_{2}$ and E$_{3}$ are well consistent with the former case. Furthermore, \citet{Archontis:2008} showed in their 3D numerical simulations of an emerging magnetic flux rope that the successful eruption of the flux rope results from the effective reconnection between the ambient field overlaying the rope and pre-existing coronal field: i.e., the downward tension force of the ambient field above the flux rope is reduced via the reconnection so that the rope can erupt. We therefore conclude that reconnection-favored magnetic fields in the flaring emerging flux regions $E_{2}$ and $E_{3}$ play a crucial role in producing the multiple flare-associated CME events in the active region NOAA 9236. As a future work, more flare-associated CMEs will be examined by using high spatial and temporal resolution data observed by the Solar Dynamics Observatory \citep[$SDO$,][]{Pesnell:2012} to better understand the long-term precondition and onset mechanism of reconnection-driven CMEs.

\acknowledgments The authors thank the anonymous referee for highly detailed and valuable comments to improve the level of this paper. We thank the teams of SOHO/EIT, SOHO/LASCO, and SOHO/MDI for providing the valuable data. SOHO is a project of international cooperation between ESA and NASA. We acknowledge use of the CME catalog generated and maintained at the CDAW Data Center by NASA and The Catholic University of America in cooperation with the Naval Research Laboratory. This research has made use of NASA's Astrophysics Data System Abstract Service. This work has been support by the ``Development of Korea Space Weather Prediction Center" project of KASI and the KASI basic research fund. This work was partially supported by a Grants-in-Aid for Scientific Research (B) ``Understanding and Prediction of Triggering Solar Flares" (23340045, Head Investigator: K. Kusano) from the Ministry of Education, Science, Sports, Technology, and Culture of Japan.

\clearpage


\begin{deluxetable}{ccccrccccc}
\tablecolumns{1}
\tabletypesize{\scriptsize}
\tablewidth{0pt}
\tablecaption{List of Eight Flare-associated CME Events \label{sec2:tab1}}
\tablehead{\multirow{3}{*}{No.} & \multicolumn{4}{c}{CME} & \multicolumn{2}{c}{Related Flare} \\
& \multirow{2}{*}{$t_a$\tablenotemark{a}} & \colhead{PA/AW\tablenotemark{b}} & \colhead{$v$\tablenotemark{c}} & \colhead{$a$\tablenotemark{d}} &
\multirow{2}{*}{$t_s$\tablenotemark{e}} & \multirow{2}{*}{$t_p$\tablenotemark{f}} & \multirow{2}{*}{$t_e$\tablenotemark{g}} & GOES \\
& & (deg) & (km s$^{-1}$) & (m s${}^{-2}$) & & & & Class }
\startdata %
1  & 2000-11-23 23:54 & 336/157  & 690  & 1.8      & 23:18 & 23:28 & 23:37 & M1.0 \\
2  & 2000-11-24 05:30 & Halo/360 & 1289 & 2.1      & 04:55 & 05:02 & 05:08 & X2.0 \\
3  & 2000-11-24 15:30 & Halo/360 & 1245 & $-$3.3   & 14:51 & 15:13 & 15:21 & X2.3 \\
4  & 2000-11-24 22:06 & Halo/360 & 1005 & $-$0.8   & 21:43 & 21:59 & 22:12 & X1.8 \\
5  & 2000-11-25 09:30 & Halo/360 & 675  & $-$4.7   & 09:06 & 09:20 & 09:40 & M3.5 \\
6  & 2000-11-25 19:31 & Halo/360 & 671  & $-$10.8  & 18:33 & 18:44 & 18:55 & X1.9 \\
7  & 2000-11-26 03:30 & 259/118  & 495  & $-$22.9* & 02:47 & 03:08 & 03:20 & M2.2 \\
8  & 2000-11-26 17:06 & Halo/360 & 980  & 5.8      & 16:34 & 16:48 & 16:56 & X4.0 \\
\hline
\enddata
\tablecomments{}
\tablenotetext{a}{~First appearance time in the LASCO/C2 FOV.}
\tablenotetext{b}{~Position Angle/Angular Width given by the LASCO CME catalog.}
\tablenotetext{c}{~Linear speed given by the LASCO CME catalog.}
\tablenotetext{d}{~Acceleration given by the LASCO CME catalog. The superscript `*' indicates that acceleration is uncertain due to either poor height measurement or a small number of height-time measurements.}
\tablenotetext{e}{~GOES soft X-ray flare start time.}
\tablenotetext{f}{~GOES soft X-ray flare peak time.}
\tablenotetext{g}{~GOES soft X-ray flare end time.}
\end{deluxetable}

\begin{deluxetable}{crrrr}
\tablecolumns{1}
\tabletypesize{\scriptsize}
\tablewidth{0pt}
\tablecaption{Magnetic Parameters \label{sec3:tab1}}
\tablehead{\multirow{2}{*}{Region} & $\dot{H}_{\mathrm{avg}}$\tablenotemark{a} & $\Delta H_{\mathrm{final}}$\tablenotemark{b} & $\Phi_{\mathrm{avg}}$\tablenotemark{c} & $\dot{\Phi}_{\mathrm{avg}}$\tablenotemark{d} \\
& (10$^{40}\,$Mx$^{2}\,$hr$^{-1}$) & (10$^{40}\,$Mx$^{2}$) & (10$^{20}\,$Mx) & (10$^{19}\,$Mx$\,$hr$^{-1}$)}
\startdata %
Entire           & $-$1.0 & $-$70  & 660 & 23.1 \\
R$^{+}\&$R$^{-}$ & 4.7  & 390  & 160 & $-$1.6 \\ 
E$_{1}$          & $-$3.0 & $-$250 & 40  & 1.8 \\
E$_{2}$          & $-$0.8 & $-$70  & 10  & 1.1 \\
E$_{3}$          & $-$2.1 & $-$170 & 50  & 10.9 \\
E$_{4}$          & 4.0  & 330  & 40  & 7.0 \\
\enddata
\tablecomments{All of the above magnetic parameters were calculated from the time profiles of $\dot{H}$, $\Delta H$, and $\Phi$ during the entire time period of 2000 November 23, 11:12 UT to 2000 November 26, 20:47 UT.}
\tablenotetext{a}{~Average helicity injection rate.}
\tablenotetext{b}{~Accumulated amount of helicity injection at the end of the time period.}
\tablenotetext{c}{~Average unsigned magnetic flux.}
\tablenotetext{d}{~Average rate of change of unsigned magnetic flux.}
\end{deluxetable}

\clearpage



\clearpage

\clearpage

\clearpage

\clearpage

\clearpage

\clearpage

\begin{thebibliography}{}
\bibitem[Antiochos et al.(1999)]{Antiochos:1999} Antiochos, S.~K., DeVore, C.~R., \& Klimchuk, J.~A.\ 1999, \apj, 510, 485
\bibitem[Archontis \& T{\"o}r{\"o}k(2008)]{Archontis:2008} Archontis, V., \& T{\"o}r{\"o}k, T.\ 2008, \aap, 492, L35 
\bibitem[Bateman(1978)]{Bateman:1978} Bateman, G.\ 1978, MHD Instability (Cambridge, MA: MIT Press)
\bibitem[Berger \& Field(1984)]{Berger:1984} Berger, M.~A., \& Field, G.~B.\ 1984, Journal of Fluid Mechanics, 147, 133
\bibitem[Brueckner et al.(1995)]{Brueckner:1995} Brueckner, G.~E., Howard, R.~A., Koomen, M.~J., et al.\ 1995, \solphys, 162, 357 
\bibitem[Chae(2007)]{Chae:2007} Chae, J.\ 2007, Advances in Space Research, 39, 1700
\bibitem[Chen \& Shibata(2000)]{Chen:2000} Chen, P.~F., \& Shibata, K.\ 2000, \apj, 545, 524
\bibitem[Delaboudini{\`e}re et al.(1995)]{Delaboudini:1995} Delaboudini{\`e}re, J.-P., Artzner, G.~E., Brunaud, J., et al.\ 1995, \solphys, 162, 291
\bibitem[D{\'e}moulin \& Pariat(2009)]{Demoulin:2009} D{\'e}moulin, P., \& Pariat, E.\ 2009, Advances in Space Research, 43, 1013
\bibitem[Falconer et al.(2002)]{Falconer:2002} Falconer, D.~A., Moore, R.~L., \& Gary, G.~A.\ 2002, \apj, 569, 1016
\bibitem[Falconer et al.(2006)]{Falconer:2006} Falconer, D.~A., Moore, R.~L., \& Gary, G.~A.\ 2006, \apj, 644, 1258
\bibitem[Fan \& Gibson(2004)]{Fan:2004} Fan, Y., \& Gibson, S.~E.\ 2004, \apj, 609, 1123
\bibitem[Feynman \& Martin(1995)]{Feynman:1995} Feynman, J., \& Martin, S.~F.\ 1995, \jgr, 100, 3355
\bibitem[Finn \& Antonsen(1985)]{Finn:1985} Finn, J. M. \& Antonsen, T. M.\ 1985, Comments Plasma Phys. Controlled Fusion, 9, 111
\bibitem[Gopalswamy(2006)]{Gopalswamy:2006} Gopalswamy, N.\ 2006, Journal of Astrophysics and Astronomy, 27, 243
\bibitem[Gopalswamy et al.(2009)]{Gopalswamy:2009} Gopalswamy, N., Yashiro, S., Michalek, G., et al.\ 2009, Earth Moon and Planets, 104, 295
\bibitem[Hood \& Priest(1981)]{Hood:1981} Hood, A.~W., \& Priest, E.~R.\ 1981, Geophysical and Astrophysical Fluid Dynamics, 17, 29
\bibitem[Isobe et al.(2002)]{Isobe:2002} Isobe, H., Yokoyama, T., Shimojo, M., et al.\ 2002, \apj, 566, 528
\bibitem[Jing et al.(2004)]{Jing:2004} Jing, J., Yurchyshyn, V.~B., Yang, G., Xu, Y., \& Wang, H.\ 2004, \apj, 614, 1054
\bibitem[Jing et al.(2012)]{Jing:2012} Jing, J., Park, S.-H., Liu, C., et al.\ 2012, \apjl, 752, L9
\bibitem[Kahler(1977)]{Kahler:1977} Kahler, S.\ 1977, \apj, 214, 891 
\bibitem[Kliem \& T{\"o}r{\"o}k(2006)]{Kliem:2006} Kliem, B., \& T{\"o}r{\"o}k, T.\ 2006, Physical Review Letters, 96, 255002
\bibitem[Kumar et al.(2010)]{Kumar:2010} Kumar, P., Manoharan, 
P.~K., \& Uddin, W.\ 2010, \apj, 710, 1195
\bibitem[Kumar et al.(2013)]{Kumar:2013} Kumar, P., Park, S.-H., Cho, K.-S., \& Bong, S.-C.\ 2013, \solphys, 282, 503
\bibitem[Kusano et al.(2003a)]{Kusano:2003a} Kusano, K., Maeshiro, T., Yokoyama, T., \& Sakurai, T.\ 2003a, Advances in Space Research, 32, 1917 
\bibitem[Kusano et al.(2003b)]{Kusano:2003b} Kusano, K., Yokoyama, T., Maeshiro, T., \& Sakurai, T.\ 2003b, Advances in Space Research, 32, 1931
\bibitem[Kusano et al.(2004)]{Kusano:2004} Kusano, K., Maeshiro, T., Yokoyama, T., \& Sakurai, T.\ 2004, \apj, 610, 537 
\bibitem[Kusano et al.(2012)]{Kusano:2012} Kusano, K., Bamba, Y., Yamamoto, T.~T., et al.\ 2012, \apj, 760, 31
\bibitem[LaBonte et al.(2007)]{LaBonte:2007} LaBonte, B.~J., Georgoulis, M.~K., \& Rust, D.~M.\ 2007, \apj, 671, 955
\bibitem[Linker et al.(2001)]{Linker:2001} Linker, J.~A., Lionello, R., Miki{\'c}, Z., \& Amari, T.\ 2001, \jgr, 106, 25165
\bibitem[Liu \& Schuck(2012)]{Liu:2012} Liu, Y., \& Schuck, P.~W.\ 2012, \apj, 761, 105
\bibitem[Luoni et al.(2011)]{Luoni:2011} Luoni, M.~L., D{\'e}moulin, P., Mandrini, C.~H., \& van Driel-Gesztelyi, L.\ 2011, \solphys, 270, 45
\bibitem[Manchester(2008)]{Manchester:2008} Manchester, W.\ 2008, Subsurface and Atmospheric Influences on Solar Activity, 383, 91 
\bibitem[Mikic et al.(1988)]{Mikic:1988} Mikic, Z., Barnes, D.~C., \& Schnack, D.~D.\ 1988, \apj, 328, 830
\bibitem[Nitta \& Hudson(2001)]{Nitta:2001} Nitta, N.~V., \& Hudson, H.~S.\ 2001, \grl, 28, 3801
\bibitem[Pariat et al.(2005)]{Pariat:2005} Pariat, E., D{\'e}moulin, P., \& Berger, M.~A.\ 2005, \aap, 439, 1191
\bibitem[Park et al.(2008)]{Park:2008} Park, S.-H., Lee, J., Choe, G.~S., Chae, J., Jeong, H., Yang, G., Jing, J., \& Wang, H.\ 2008, \apj, 686, 1397
\bibitem[Park et al.(2010a)]{Park:2010a} Park, S.-H., Chae, J., \& Wang, H.\ 2010a, \apj, 718, 43
\bibitem[Park et al.(2010b)]{Park:2010b} Park, S.-H., Chae, J., Jing, J., Tan, C., \& Wang, H.\ 2010b, \apj, 720, 1102 
\bibitem[Park et al.(2012)]{Park:2012} Park, S.-H., Cho, K.-S., Bong, S.-C., et al.\ 2012, \apj, 750, 48
\bibitem[Pesnell et al.(2012)]{Pesnell:2012} Pesnell, W.~D., Thompson, B.~J., \& Chamberlin, P.~C.\ 2012, \solphys, 275, 3 
\bibitem[Pevtsov(2008)]{Pevtsov:2008} Pevtsov, A.~A.\ 2008, Journal of Astrophysics and Astronomy, 29, 49
\bibitem[Romano et al.(2003)]{Romano:2003} Romano, P., Contarino, L., \& Zuccarello, F.\ 2003, \solphys, 218, 137
\bibitem[Romano et al.(2011)]{Romano:2011a} Romano, P., Pariat, E., Sicari, M., \& Zuccarello, F.\ 2011, \aap, 525, A13
\bibitem[Romano \& Zuccarello(2011)]{Romano:2011b} Romano, P., \& Zuccarello, F.\ 2011, \aap, 535, A1 
\bibitem[Scherrer et al.(1995)]{Scherrer:1995} Scherrer, P.~H., Bogart, R.~S., Bush, R.~I., et al.\ 1995, \solphys, 162, 129
\bibitem[Schuck(2006)]{Schuck:2006} Schuck, P.~W.\ 2006, \apj, 646, 1358 
\bibitem[Shibata(1999)]{Shibata:1999} Shibata, K.\ 1999, \apss, 264, 129 
\bibitem[Smyrli et al.(2010)]{Smyrli:2010} Smyrli, A., Zuccarello, F., Romano, P., et al.\ 2010, \aap, 521, A56  
\bibitem[Takasaki et al.(2004)]{Takasaki:2004} Takasaki, H., Asai, A., Kiyohara, J., et al.\ 2004, \apj, 613, 592
\bibitem[Tan et al.(2009)]{Tan:2009} Tan, C., Chen, P.~F., Abramenko, V., \& Wang, H.\ 2009, \apj, 690, 1820
\bibitem[Titov \& D{\'e}moulin(1999)]{Titov:1999} Titov, V.~S., \& D{\'e}moulin, P.\ 1999, \aap, 351, 707
\bibitem[T{\"o}r{\"o}k \& Kliem(2005)]{Torok:2005} T{\"o}r{\"o}k, T., \& Kliem, B.\ 2005, \apjl, 630, L97
\bibitem[T{\"o}r{\"o}k et al.(2013)]{Torok:2013} T{\"o}r{\"o}k, 
T., Temmer, M., Valori, G., et al.\ 2013, \solphys, 70
\bibitem[Tsuneta et al.(1991)]{Tsuneta:1991} Tsuneta, S., Acton, L., Bruner, M., et al.\ 1991, \solphys, 136, 37
\bibitem[van Ballegooijen \& Martens(1989)]{vanBallegooijen:1989} van Ballegooijen, A.~A., \& Martens, P.~C.~H.\ 1989, \apj, 343, 971
\bibitem[Wang et al.(1998)]{Wang:1998} Wang, H., Denker, C., Spirock, T., et al.\ 1998, \solphys, 183, 1
\bibitem[Wang et al.(2002)]{Wang:2002} Wang, H., Gallagher, P., Yurchyshyn, V., Yang, G., \& Goode, P.~R.\ 2002, \apj, 569, 1026  
\bibitem[Wang et al.(2004)]{Wang:2004} Wang, J., Zhou, G., \& Zhang, J.\ 2004, \apj, 615, 1021
\bibitem[Yamamoto et al.(2005)]{Yamamoto:2005} Yamamoto, T.~T., Kusano, K., Maeshiro, T., Yokoyama, T., \& Sakurai, T.\ 2005, \apj, 624, 1072 
\bibitem[Yan et al.(2012)]{Yan:2012} Yan, X.~L., Qu, Z.~Q., 
Kong, D.~F., \& Xu, C.~L.\ 2012, \apj, 754, 16
\bibitem[Yang et al.(2004)]{Yang:2004} Yang, G., Xu, Y., Cao, W., et al.\ 2004, \apjl, 617, L151
\bibitem[Yashiro et al.(2004)]{Yashiro:2004} Yashiro, S., Gopalswamy, N., Michalek, G., et al.\ 2004, Journal of Geophysical Research (Space Physics), 109, 7105 
\bibitem[Yokoyama et al.(2003)]{Yokoyama:2003} Yokoyama, T., Kusano, K., Maeshiro, T., \& Sakurai, T.\ 2003, Advances in Space Research, 32, 1949 
\bibitem[Zhang \& Wang(2002)]{Zhang:2002} Zhang, J., \& Wang, J.\ 2002, \apjl, 566, L117 
\bibitem[Zuccarello et al.(2011)]{Zuccarello:2011} Zuccarello, F.~P., Romano, P., Zuccarello, F., \& Poedts, S.\ 2011, \aap, 530, A36 
\end{thebibliography}
\end{document}